\documentclass{PoS}
\usepackage{graphicx}
\usepackage{amssymb}
\usepackage{amsmath}
\usepackage{bm}
\usepackage{slashed}
\usepackage{datetime}
\usepackage{mciteplus}
\usepackage{multirow}
\usepackage{color}
\usepackage[utf8]{inputenc}
\usepackage[normalem]{ulem}
\usepackage{float} 
\usepackage[caption=false]{subfig} 

\newcommand{\smarrow}{\mbox{\raisebox{-4.5pt}[0pt][0pt]{$\hspace{-1pt} 
		\vec{\phantom{v}}$}}}
\newcommand{\hermes}{\textsc{Hermes}}
\newcommand{\compass}{\textsc{Compass}}

\newcommand{\T}{\perp}
\newcommand{\Tperp}{T}
\newcommand{\bT}{\xi_T}

\title{Extraction of partonic transverse momentum distributions from semi-inclusive deep inelastic scattering and Drell-Yan data}

\ShortTitle{Extraction of partonic transverse momentum distributions from SIDIS  and DY data}

\author{\speaker{Cristian Pisano}\\
        Dipartimento di Fisica, Universit\`a di Pavia and INFN, Sezione di Pavia \\
        Via Bassi 6, I-27100 Pavia, Italy\\
        E-mail: \email{cristian.pisano@unipv.it}}

\author{Alessandro Bacchetta\\
         Dipartimento di Fisica, Universit\`a di Pavia and INFN, Sezione di Pavia \\
        Via Bassi 6, I-27100 Pavia, Italy\\
        E-mail: \email{alessandro.bacchetta@unipv.it}}

\author{Filippo Delcarro\\
        Dipartimento di Fisica, Universit\`a di Pavia and INFN, Sezione di Pavia \\
        Via Bassi 6, I-27100 Pavia, Italy\\
        E-mail: \email{filippo.delcarro@pv.infn.it}}

\author{Marco Radici\\
        INFN, Sezione di Pavia \\
        Via Bassi 6, I-27100 Pavia, Italy\\
        E-mail: \email{marco.radici@pv.infn.it}}

\author{Andrea Signori\\
        Theory Center, Thomas Jefferson National Accelerator Facility\\
        12000 Jefferson Avenue, Newport News, VA 23606, USA\\
        E-mail: \email{asignori@jlab.org}}

\abstract{We present a first attempt at a global fit of unpolarized quark transverse momentum dependent distribution and fragmentation functions from available data on semi-inclusive deep-inelastic scattering, Drell-Yan and $Z$ boson production processes. This analysis is performed in the low transverse momentum region, at leading order in perturbative QCD and with the inclusion of energy scale evolution effects at the next-to-leading logarithmic accuracy.}

\FullConference{XXV International Workshop on Deep-Inelastic Scattering and Related Subjects\\
		3-7 April 2017\\
		University of Birmingham, UK}

\begin{document}

\section{Introduction and formalism}

Transverse momentum dependent parton distribution and fragmentation functions (TMDs) encode fundamental information on the intrinsic motion of partons and the correlation between their spins and momenta, providing a full three-dimensional picture of hadrons in momentum space. Similarly to the more common collinear parton distributions, TMDs are not purely perturbative quantities. As such, they cannot be derived entirely from first principles: only their energy scale dependence can be calculated within the framework of perturbative QCD. Their full determination therefore requires an experimental input as well. 

In this contribution to the proceedings, based on \cite{Bacchetta:2017gcc} to which we refer for details, we describe the extraction of  unpolarized quark TMDs from semi-inclusive deep-inelastic scattering (SIDIS), Drell-Yan (DY) and $Z$ boson production data, coming from different experiments. In this way, we are able not only to gather information on the intrinsic transverse momenta of the TMDs, but also to study their evolution over a large energy range, as well as  to test their universality properties among different processes.

Most of the analyzed data refer to SIDIS hadron multiplicities, which are defined as  
\begin{equation}
m_N^h (x,z,|\bm{P}_{h\Tperp}|, Q^2) = \frac{d \sigma_N^h / ( dx \, dz \,d|\bm{P}_{h\Tperp}|\, dQ^2) }
                                                                   {d\sigma_{\text{DIS}} / ( dx \,dQ^2 ) }\approx \frac{2 \pi\,|\bm{P}_{h\Tperp}| F_{UU ,T}(x,z,\bm{P}_{h\Tperp}^2, Q^2)}{F_{T}(x,Q^2) } \, ,
\label{e:multiplicity}
\end{equation}
where $d\sigma_N^h$ and $F_{UU,T}$ are, respectively,  the cross section and the transverse structure function for the SIDIS reaction $\ell(l) + N(P) \to \ell(l') + h(P_h) + X$. Similarly, $d\sigma_{\text{DIS}}$ and $F_{T}$ are the well-known corresponding quantities in the inclusive process. We adopt the familiar DIS variables $x$, $y$ and $Q^2=-q^2 = -(l-l^\prime)^2$; $\bm P_{hT}$ is the component of final hadron three-momentum $\bm P_h$  transverse to $\bm {q}$, while $z = {P \cdot P_h}/{P\cdot q}$. The approximation in (\ref{e:multiplicity}) is only valid in the kinematic region under study, defined by the constraints $\bm{P}_{hT}^2 \ll Q^2$ and  $M^2 \ll Q^2$, with $M$ being the mass of the nucleon $N$. 

In order to consider TMD evolution, we introduce the Fourier transforms of the distribution  $f_1^a\big(x, \bm{k}_\T^2;Q^2\big)$ and fragmentation function  $D_1^{a\smarrow h}\big(z, \bm{P}_{\T}^2; Q^2 \big)$ for a quark with flavor $a$ and electric charge $e_a$, namely
\begin{align} 
\tilde{f}_1^a\big(x, \bT^2;Q^2\big) &=
\int_0^{\infty} d |\bm{k}_\T| 
                |\bm{k}_\T|J_0\big(\bT |\bm{k}_\T|\big) 
       f_1^a\big(x, \bm{k}_\T^2;Q^2\big)\,,\\
\tilde{D}_1^{a\smarrow h}\big(z, \bT^2; Q^2 \big) &=
\int_0^{\infty} \frac{d |\bm{P}_{\T}|}{z^2} |\bm{P}_{\T}| 
                                             J_0\big(\bT |\bm{P}_{\T}|/z\big)
       D_1^{a\smarrow h}\big(z, \bm{P}_{\T}^2; Q^2 \big)~.
\end{align}  
Hence the structure function $F_{UU,T}$, at leading order (LO) in perturbative QCD, takes the form
\begin{align}
\label{e:SIDISkTFF}
   F_{UU,T}(x,z, \bm{P}_{h \Tperp}^2, Q^2) &\approx 2\pi \sum_a e_a^2 x 
       \int_0^{\infty} {d \bT} \bT J_0\big(\bT |\bm{P}_{hT}|/z\big)
      \tilde{f}_1^a\big(x, \bT^2;Q^2\big) \tilde{D}_1^{a\smarrow h}\big(z, \bT^2;
      Q^2 \big)~. 
\end{align} 
Analogous expressions for the DY and $Z$ production cross sections can be found in \cite{Bacchetta:2017gcc}.

\begin{figure}[t]
\begin{center}
\hspace*{-1.7cm}
\includegraphics[trim = 0cm 10.2cm 0cm 4.6cm , clip,width=1.25\textwidth]{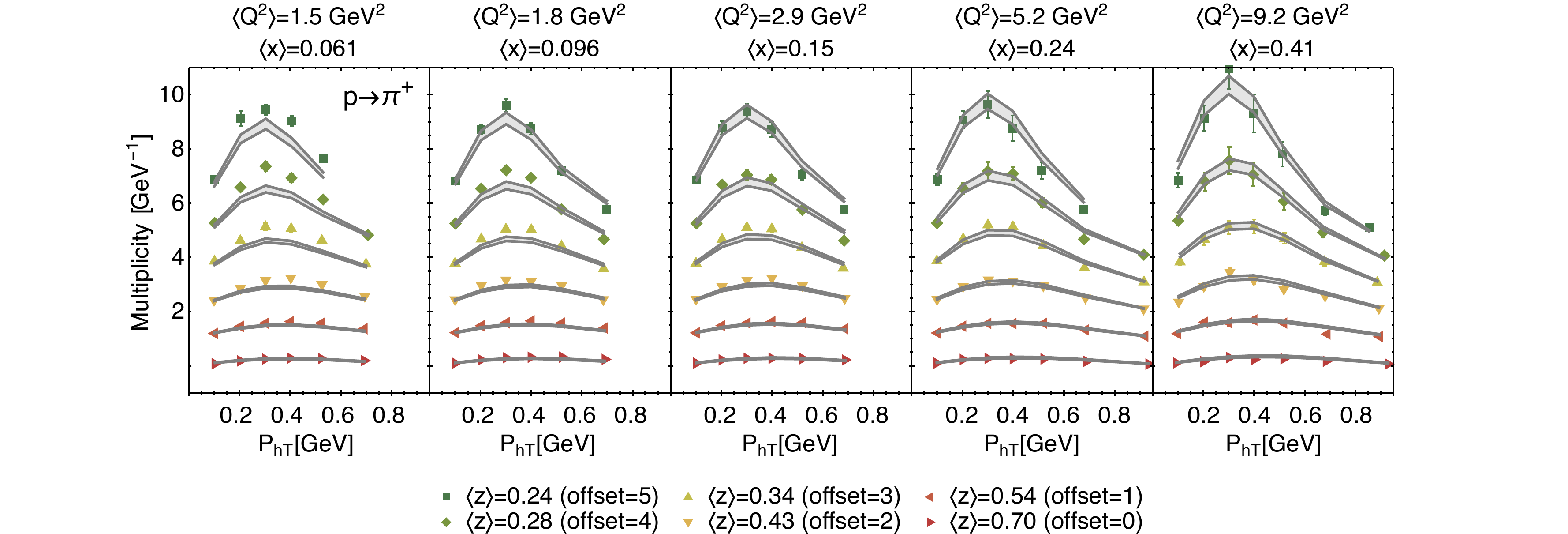}
\vspace{-0.7cm}
\end{center}
\caption{\hermes\ multiplicities for the process $ep\to e \pi^+X$  as a function of the transverse momentum of the detected hadron  $P_{hT}$ at different $\langle x \rangle$ , $\langle z \rangle$, $\langle Q^2 \rangle$ bins.  Each $\langle z \rangle$  bin has been shifted for clarity by an offset.} 
\label{f:H_pions}
\end{figure}

\section{TMD evolution}
The evolved TMDs at LO in configuration space are given by~\cite{Bacchetta:2017gcc,Collins:2011zzd}
\begin{align}   
\widetilde{f}_1^a (x,  \bT^2; Q^2) &= f_1^a (x ; \mu_b^2) 
\  e^{S (\mu_b^2, Q^2)} \  e^{g_K(\bT) \ln (Q^2 / Q_0^2)} \  \widetilde{f}_{1 {\rm NP}}^a (x, \bT^2) \ ,
\label{e:TMDevol1b} \\
\widetilde{D}_1^{a\to h} (z, \bT^2; Q^2) &= D_1^{a\to h} (z; \mu_b^2) \  e^{S (\mu_b^2, Q^2)} \  e^{g_K( \bT) \ln (Q^2 / Q_0^2)} \  \widetilde{D}_{1 {\rm NP}}^{a\to h} (z, \bT^2) \  ,
\label{e:TMDevol2b}
\end{align}
where $f_1^a (x ; \mu_b^2) $ and $D_1^{a\to h} (z; \mu_b^2) $ are the usual collinear distribution and fragmentation functions, evaluated at the initial energy scale $\mu_b$, which is chosen such that at $Q_0=1$ GeV there are no evolution effects~\cite{Bacchetta:2017gcc}.  The general expression of the Sudakov exponent $S$ reads\footnote{Notice that a factor $1/2$ is missing in the definition of the Sudakov exponent given in (2.3) of \cite{Bacchetta:2017gcc}.}
\begin{equation} 
S(\mu_b^2,Q^2)=- \frac{1}{2}\int_{\mu_b^2}^{Q^2}{d\mu^2\over \mu^2}
\bigg[A\Big(\alpha_S(\mu^2)\Big)\ln\bigg({Q^2\over \mu^2}\bigg) 
+ B\Big(\alpha_S(\mu^2)\Big) \bigg] \ ,
\label{e:Sudakov} 
\end{equation} 
where, to the next-to-leading logarithmic (NLL) accuracy, $A$ and $B$ are expanded as
\begin{align}
A &= C_F\bigg(\frac{\alpha_S}{\pi} \bigg) +  \frac{1}{2}\,C_F
\bigg(\frac{\alpha_S}{\pi} \bigg)^2 \bigg[
C_A \bigg( \frac{67}{18} - \frac{\pi^2}{6} \bigg)
- \frac{5}{9} N_f \bigg],
&
B &= - \frac{3}{2}C_F
\bigg(\frac{\alpha_S}{\pi} \bigg).
\end{align} 
For the nonperturbative Sudakov factor in (\ref{e:TMDevol1b}) and (\ref{e:TMDevol2b}), we choose  $g_K (\bT) = - g_2 \bT^2 / 2$,
with $g_2$ being a free parameter. Finally, the nonperturbative parts of the TMDs are parametrized as follows
\begin{align}
\widetilde{f}_{1 {\rm NP}}^a (x, \bT^2) &= \frac{1}{2\pi}
        e^{-g_{1a} \frac{\bT^2}{4}}
        \bigg( 1 - \frac{\lambda  g_{1a}^{2}}{1+\lambda g_{1a}}  \frac{\bT^2}{4} \bigg)\  ,
\label{e:f1NP} \\
\widetilde{D}_{1 {\rm NP}}^{a \to h} (z, \bT^2) &= 
    \frac{ g_{3 a\to h} \   e^{-g_{3 a\to h} \frac{\bT^2}{4 z^2}}
        + \big(\lambda_F/z^2\big)    g_{4 a\to h}^{2}
    \left(1 - g_{4 a\to h} \frac{\bT^2}{4 z^2} \right)
         \  e^{- g_{4 a\to h} \frac{\bT^2}{4z^2}}}
     {2 \pi z^2 \Big(g_{3 a\to h} + \big(\lambda_F/z^2\big)    g_{4 a\to h}^{2}\Big)} \,,
\label{e:D1NP}
\end{align} 
\begin{figure}[t]
\centering
\includegraphics[trim = 0cm 0.3cm 0cm 1.9cm , clip,width=1.\textwidth]{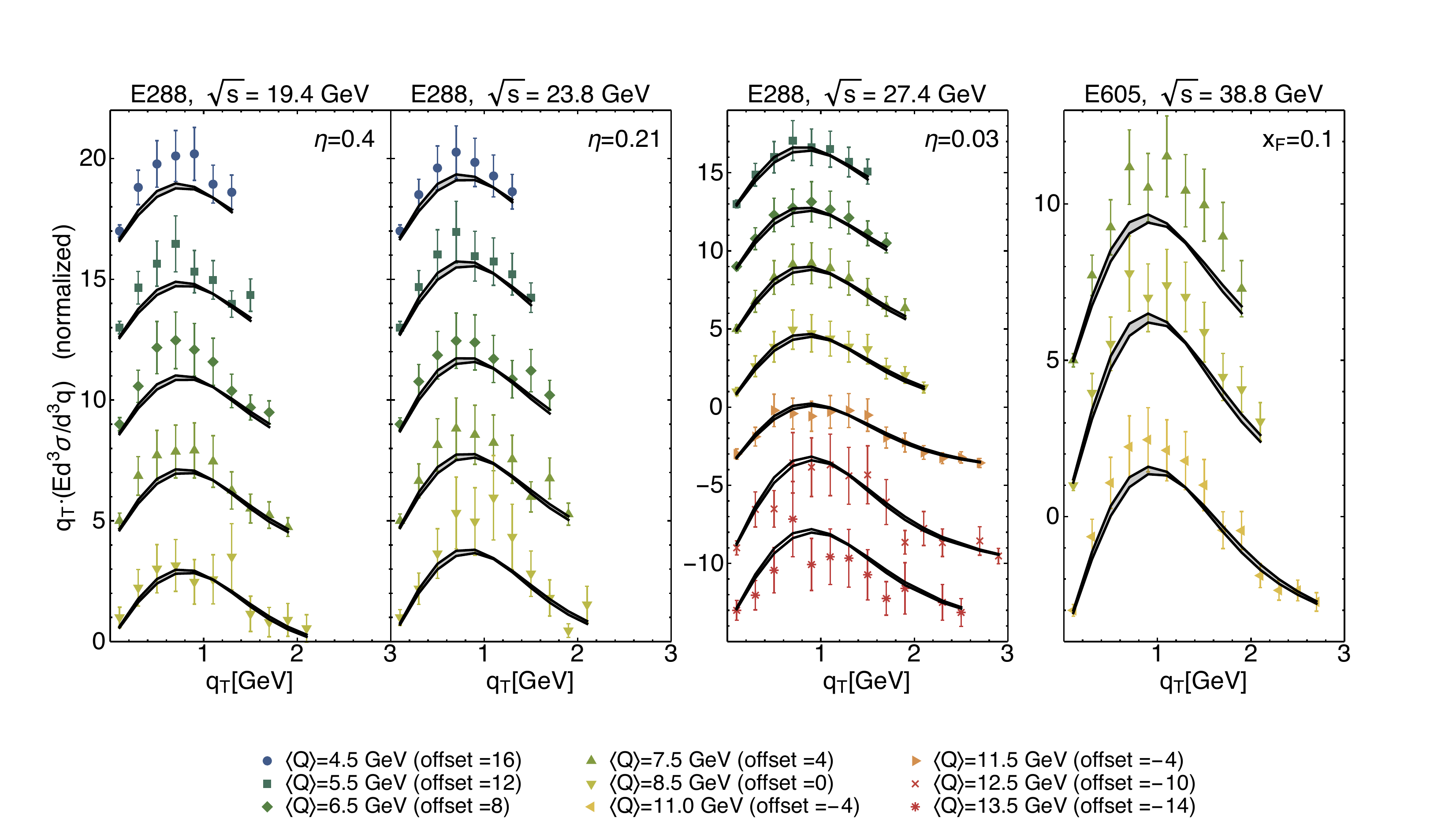}
\caption{DY cross section as a function of the $q_T$ of the virtual photon for different values of $\sqrt{s}$ and  $\langle Q \rangle$. For clarity, each $\langle Q \rangle$ bin has been normalized 
  and then shifted by an
  offset as indicated in the legend.}
\label{f:DY_panel}
\end{figure}
\begin{figure}[b]
\begin{center}
\includegraphics[trim = 0cm 1.3cm 0cm 2cm , clip,width=0.9\textwidth]{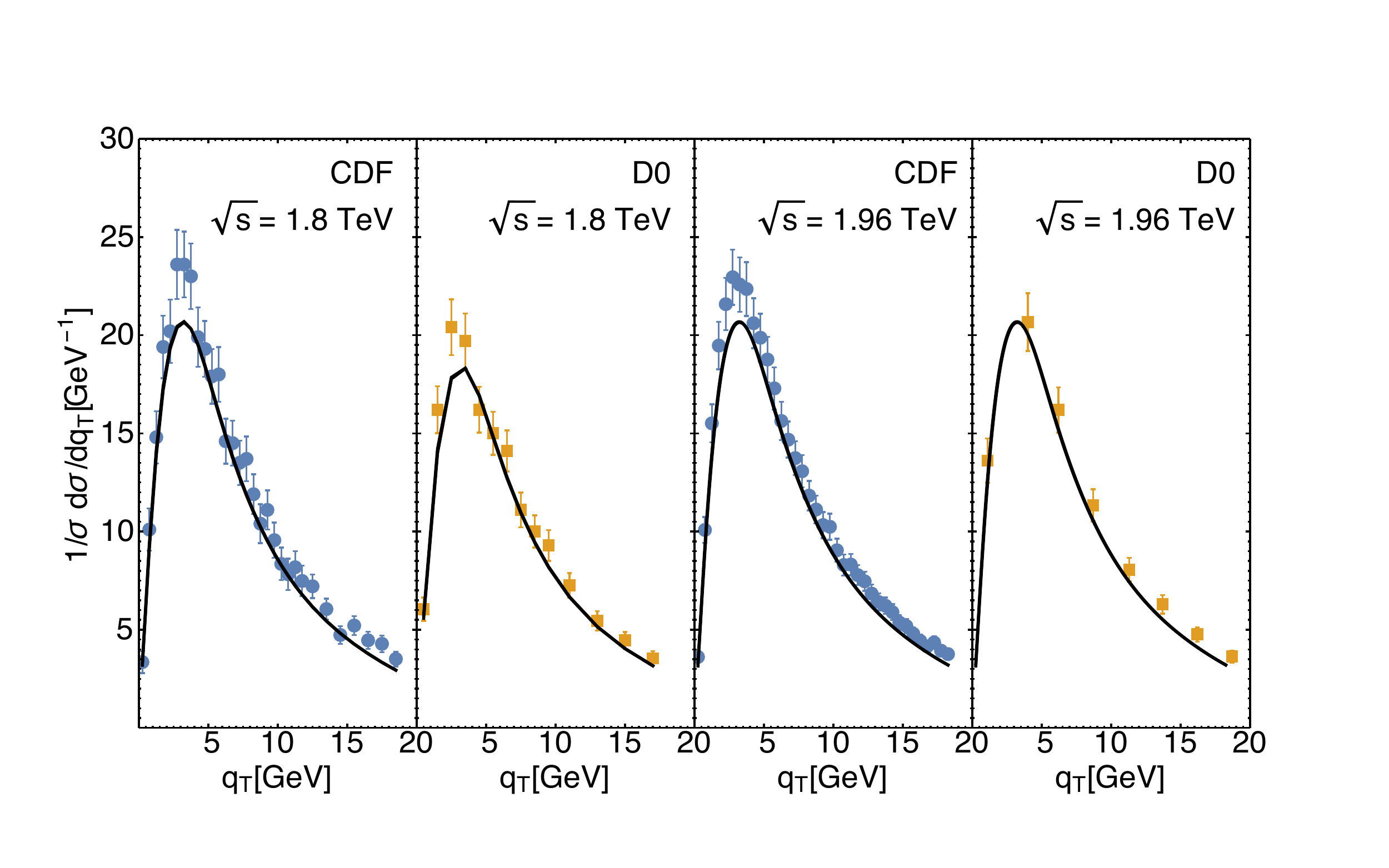}
\end{center}
\caption{Cross section for the process $p\bar p \to Z\, X$ as a function of  the transverse momentum $q_T$ of the $Z$ boson, for  different energies of the  CDF and D0 experiments at the Tevatron.} 
\label{f:Z_qT}
\end{figure}
The Gaussian width $g_1$ of the distribution function is assumed to depend on the fractional longitudinal momentum $x$ according to
\begin{equation} 
 g_1 (x) = N_1 \;  
\frac{(1-x)^{\alpha} \  x^{\sigma} }{ (1 - \hat{x})^{\alpha} \  \hat{x}^{\sigma} } \, ,
\label{e:kT2_kin}
\end{equation}
where $\hat{x} = 0.1$, while $\alpha, \, \sigma$,  $N_1 \equiv  g_1 (\hat{x})$, together with $\lambda$ in (\ref{e:f1NP}), are free parameters.
Analogously, 
\begin{align}  
g_{3,4} (z) = N_{3,4} \  
               \frac{ (z^{\beta} + \delta)\ (1-z)^{\gamma} }{ (\hat{z}^{\beta} + \delta)\   (1 - \hat{z})^{\gamma} } \, ,
 \label{e:PT2_kin}
 \end{align}
where $\hat{z} = 0.5$, while $\beta, \, \gamma, \, \delta $, $N_{3,4} \equiv g_{3,4} (\hat{z})$, as well as $\lambda_F$ in (\ref{e:D1NP}), are the free parameters for the fragmentation functions to be fitted to the data.  The average transverse momentum squared of the TMDs at $Q = Q_0$ can therefore be expressed as
\begin{align}
\big \langle \bm{k}_{\perp}^2 \big \rangle (x) &= \frac{g_1(x) + 2 \lambda g_1^2(x)}
{1+ \lambda g_1(x)},
&
\big \langle \bm{P}_{\perp}^2 \big \rangle (z) &= \frac{g_3^2(z) + 2 \lambda_F
  g_4^3(z)}{g_3(z) + \lambda_F g_4^2(z)}~.
\label{e:transmom2}
\end{align}
In the present study any possible  flavor dependence has been neglected.

\section{Numerical results}

We consider SIDIS data off proton~\cite{Airapetian:2012ki} and deuteron targets~\cite{Airapetian:2012ki,Adolph:2013stb}, DY events at low energy~\cite{Ito:1980ev,Moreno:1990sf} and $Z$ boson production at the Tevatron~\cite{Affolder:1999jh,Abbott:1999wk,Aaltonen:2012fi,Abazov:2007ac}. We identify the current fragmentation region in SIDIS by applying the cut $0.2 < z < 0.7$. Moreover, since TMD factorization requires the presence of two separate energy scales, we impose  $Q^2 > 1.4$ GeV$^2$ and restrict our fit to the small transverse momentum region, by selecting the maximum value of the transverse momenta measured in each process on the basis of phenomenological  considerations~\cite{Bacchetta:2017gcc}. After this selection, the total number of data points is 8059. 

The unpolarized TMDs are extracted by performing a fit to the data of our 11 free parameters, using a replica methodology~\cite{Bacchetta:2017gcc,Signori:2013mda}. The  $\chi^2$/d.o.f. is quite good, its average value being $1.55 \pm 0.05$.  It can be improved down to 1.02, without changing the parameters, by 
restricting the kinematic cuts in order to  better identify the region where TMD factorization is supposed to hold.

\begin{figure}[t]
\vspace{-0.2cm}
\centering
\subfloat[]{\includegraphics[width=0.45\textwidth]{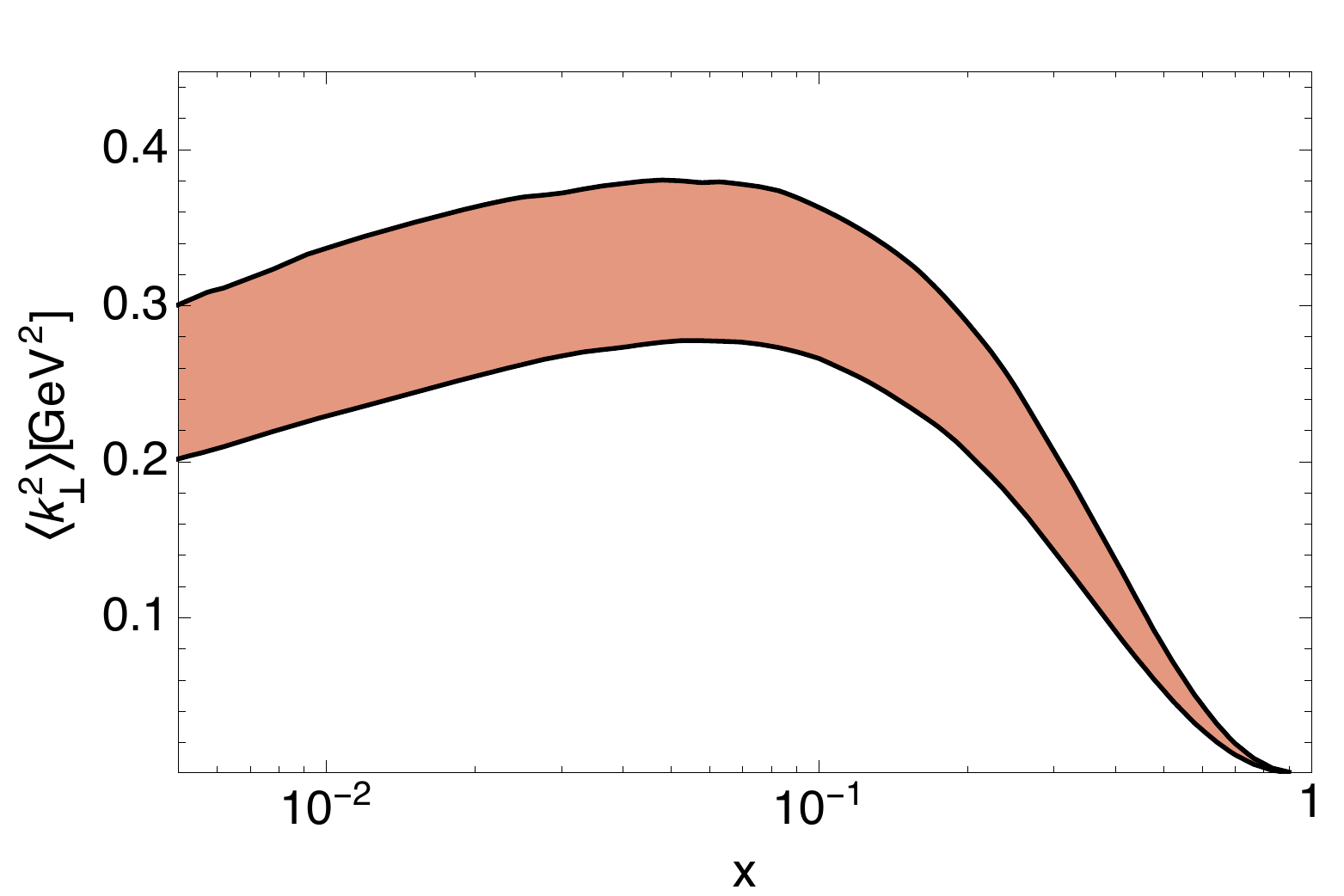}}
\hspace*{0.5cm}
\subfloat[]{\includegraphics[width=0.45\textwidth]{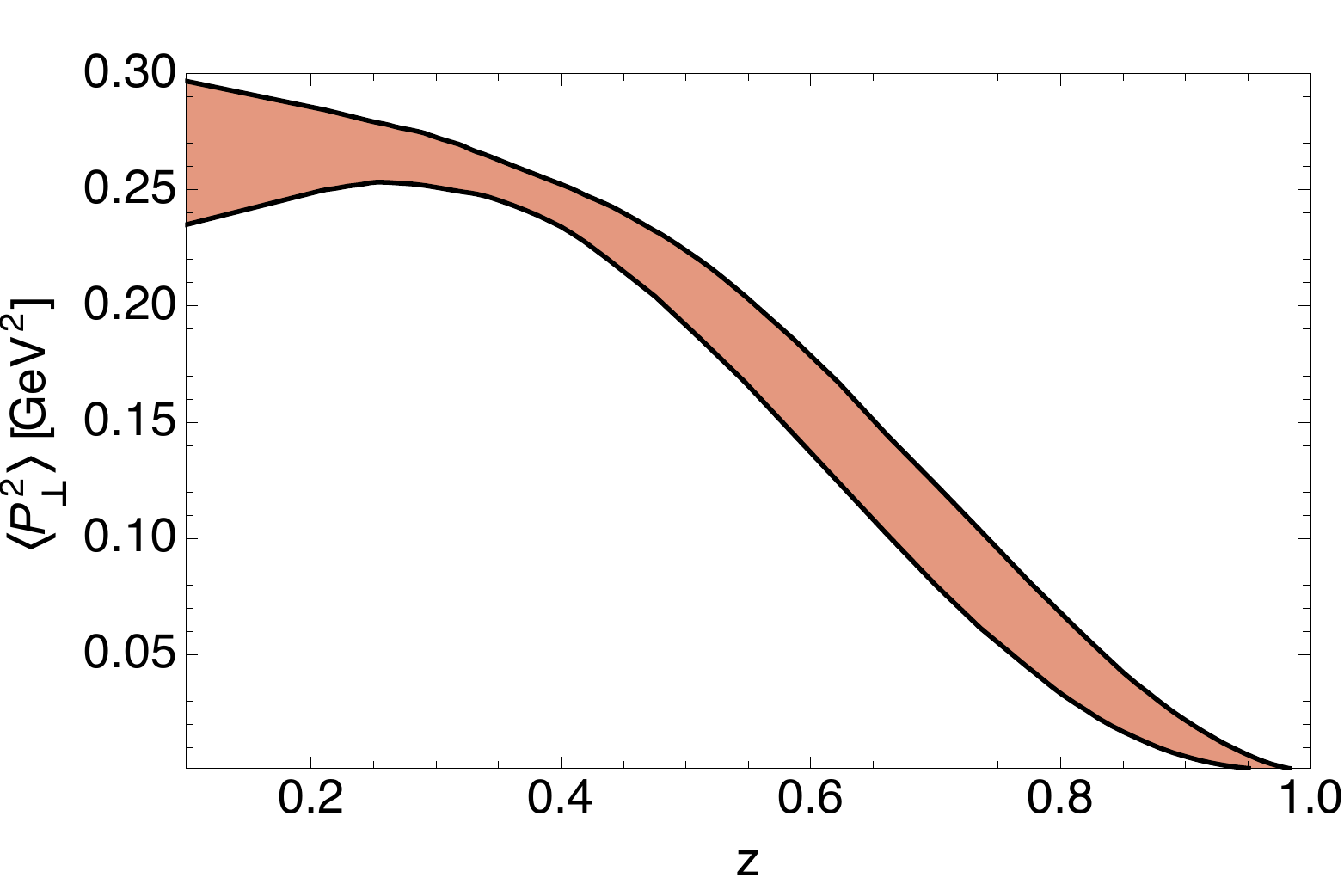}}
\caption{$\big \langle \bm{k}_{\T}^2 \big \rangle$ as a function of $x$
  (a) and $\big \langle \bm{P}_{\perp}^2 \big \rangle$ as a function of $z$ (b), both calculated at $Q^2= 1$ GeV$^2$. }
\label{f:avmomenta_68CL}
\end{figure}

For illustration, in Fig.~\ref{f:H_pions} our results are compared with  the \hermes\ multiplicities for the production of positively charged pions off a proton at different $\langle x\rangle$, $\langle z \rangle $, $\langle Q^2\rangle$ bins, as a function of  the transverse momentum of the final pion $P_{hT}$.  The bands are computed as the $68\%$ confidence level envelope of the full sets of curves from all the 200 replicas\footnote{We point out that in our analysis of \compass~measurements,  which are affected by normalization errors~\cite{Adolph:2013stb},  we fit normalized  multiplicities, obtained by dividing the data in each bin in $(x,z,Q^2,P_{hT})$  by the data point with the lowest $P_{hT}$ in the bin.}. Results from DY and Z-boson productions are presented in Figs.~\ref{f:DY_panel} and \ref{f:Z_qT}, respectively. We note that, because of TMD evolution, the position of the peak shifts from $q_T\sim 1$ GeV for DY events to $q_T\sim 7$ GeV for $Z$ production. Finally, the resulting average transverse momentum squared for the TMDs are presented in Fig.~\ref{f:avmomenta_68CL} at the scale $Q=1$ GeV, where the TMDs coincide with their nonperturbative input.  

\section{Conclusions}
We have shown for the first time that it is possible  to perform a simultaneous extraction of unpolarized TMD distributions and fragmentation functions
from SIDIS, Drell-Yan and $Z$ boson production data in the small transverse momentum region, collected in several experiments at different energies. We have found that most of the discrepancies with the measurements come from the normalization and not from the transverse momentum shape. Such a tension could probably be 
relaxed by a more precise analysis from the perturbative point view.  Moreover, in future studies, the description at low transverse momentum should be properly matched to the collinear fixed-order calculations at high transverse momentum. Further improvements could be achieved by exploring  different functional forms for all the nonperturbative ingredients, possibly including a  flavor dependence of the intrinsic transverse momenta as well.

\begin{acknowledgments}
This work is supported by the European Research Council (ERC) under the European Union's Horizon 2020 research and innovation program (grant agreement No. 647981, 3DSPIN). 
AS acknowledges support from U.S. Department of Energy contract DE-AC05-06OR23177. 
\end{acknowledgments}

\end{document}